\documentclass[final]{IEEEtran}
\usepackage[update,prepend]{epstopdf}

\usepackage{graphics}
\usepackage{multirow}
\usepackage{tikz}
\usepackage{bbm} % for \mathbbm{1}
\usepackage{pdfpages}
\usepackage{multirow}
\usepackage{subfig}
\usepackage{comment}
\usepackage{makecell}

\usepackage{setspace}	% Remove in double column version. Also search for \setstretch in the body of the paper and comment these commands for double column
\usepackage{graphicx}
\usepackage{algorithm,algorithmic}
\usepackage{multicol}

\usepackage[justification=centering]{caption}
\usepackage{textcomp}
\usepackage{psfrag}
\usepackage{arydshln}
\usepackage{url}
\usepackage{soul}
\usepackage{graphicx,color}
\usepackage[nolist]{acronym}
\usepackage{array}

\usepackage{mathtools,lipsum}
\usepackage{cuted}
\usepackage{amsmath}
\usepackage{graphicx}
\usepackage{threeparttable}

\usepackage{booktabs}
\usepackage{amssymb}   % for \checkmark

\begin{document}
%\pagenumbering{gobble}
\graphicspath{{./Figures/}}

%\title{
%High-Altitude Platform Station:
%A Key Stratospheric Player in the Nexus of Non-Terrestrial Networks
%}
\title{
Toward Environment-Aware LAE: SAR as a Shared Sensing Infrastructure
}
\author{
 
Xue Zhang, {\em Graduate Student Member, IEEE}, Bang Huang, {\em Member, IEEE}, and \\ Mohamed-Slim Alouini, {\em Fellow, IEEE}
\thanks{The authors are with King Abdullah University of Science and Technology (KAUST), CEMSE division, Thuwal 23955-6900, Saudi Arabia (e-mail: xue.zhang@kaust.edu.sa; bang.huang@kaust.edu.sa; slim.alouini@kaust.edu.sa) (Corresponding author: Bang Huang).}
\vspace{-6mm}
}
\maketitle
\thispagestyle{empty}
\begin{abstract}
The rapid growth of the low-altitude economy (LAE) is making aerial systems an important part of future digital infrastructure. Although major advances have been achieved in unmanned aerial vehicle (UAV) platforms, communications, and autonomous control, environmental perception remains a key bottleneck to reliable and scalable LAE operations. Existing sensing modalities, such as optical, LiDAR, and millimeter-wave radar, are limited by visibility, sensing range, and environmental conditions, resulting in fragmented situational awareness. This article argues that addressing these limitations requires a shift from platform-centric sensing to a shared, environment-aware sensing infrastructure. In this context, synthetic aperture radar (SAR) offers a distinct advantage by enabling all-weather, wide-area perception. We show that SAR can support UAV operations through global environmental awareness, enhance task-level sensing, and enable cooperative sensing across satellites, high-altitude platforms, UAVs, and ground systems. Building on this perspective, we present a system-level view of SAR-enabled LAE, highlighting key transformations from fragmented to infrastructure-centric sensing, from reactive to predictive operation, and from device-centric to environment-aware networking. We further discuss enabling architectures, including multi-platform sensing hierarchies, integration with integrated sensing and communication (ISAC), and the role of artificial intelligence and digital twins, along with the key challenges toward real-world deployment. By positioning SAR as a shared sensing foundation rather than a standalone modality, this article provides new insights into the design of scalable, reliable, and intelligent LAE systems.
\end{abstract}

\section{Introduction}
\label{sec_introduction}
The low-altitude economy (LAE)~\cite{he2025ubiquitous,cai2025secure} is rapidly emerging as one of the most dynamic frontiers of next-generation digital infrastructure. Driven by advances in unmanned aerial vehicles (UAVs), wireless connectivity, and intelligent automation, LAE is attracting significant attention from governments, industry, and academia worldwide. Applications such as aerial logistics, energy inspection, precision agriculture, tourism, and emergency response are no longer isolated demonstrations, but are evolving into large-scale, commercially viable services. As a result, low-altitude airspace is transforming into a highly active and economically valuable domain, comparable in importance to terrestrial transportation networks and satellite systems. This trend is further accelerated by the convergence of 6G, non-terrestrial networks (NTN), and artificial intelligence, which are reshaping how low-altitude systems are connected, coordinated, and operated.
However, the rapid expansion of LAE is also exposing a fundamental limitation. While substantial progress has been made in enabling UAV flight, communication, and control, the ability to reliably perceive and understand the low-altitude environment remains insufficient. Looking ahead, the success of LAE will not be determined solely by how well UAVs can fly, but by whether the airspace itself can be continuously sensed, interpreted, and managed. In this sense, LAE is transitioning from a flight-centric paradigm toward an environment-aware system, where large-scale missions, high-density operations, and autonomous coordination rely fundamentally on continuous and trustworthy environmental perception.

Despite this rapid progress, the sensing foundation of current LAE systems remains fundamentally limited. When hundreds or even thousands of UAVs operate in complex environments such as urban canyons, mountain valleys, coastal regions, and disaster zones, existing approaches struggle to scale~\cite{yuan2025ground}. Manual supervision quickly becomes infeasible under high traffic density, while vision-based sensing degrades sharply under fog, rain, glare, shadows, and occlusions. Other modalities, including optical cameras, LiDAR, millimeter-wave radar, and communication-assisted sensing, each operate reliably only within restricted environmental and operational conditions. 
As a result, perception failures tend to occur precisely in the most safety-critical scenarios, leading to fragmented situational awareness, unreliable decision making, and limited coordination across platforms. In effect, UAVs are often forced to operate as locally reactive agents, relying on incomplete and unstable observations of their immediate surroundings.
More importantly, this limitation is not simply due to the weakness of individual sensors, but reflects a deeper structural gap. Current sensing approaches are inherently local, platform-centric, and environment-dependent, making it fundamentally difficult to provide continuous, large-scale, and reliable awareness across the entire airspace. Consequently, LAE systems lack a global and consistent view of the environment, preventing proactive planning, coordinated operation, and large-scale system optimization.
Bridging this gap requires more than incremental improvements to existing sensing modalities. It calls for a shared sensing foundation that can provide persistent, wide-area, and environment-level awareness, transforming UAVs from locally reactive entities into globally informed and collaboratively operating agents.

\begin{figure*}[htp]
\centering
\includegraphics[width=1.0\linewidth]{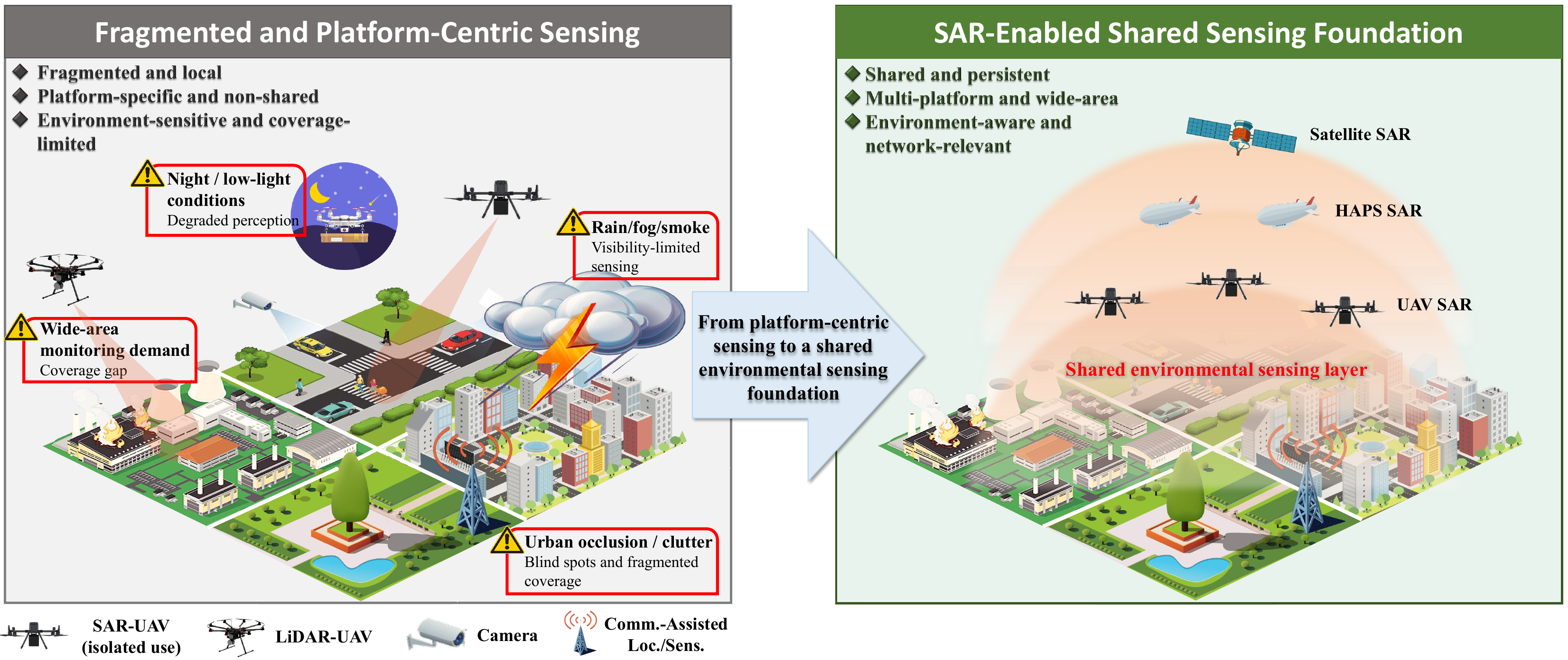}
\caption{Toward a shared sensing foundation for the LAE.}
\label{fig:overview_transition}
\end{figure*}

In this context, synthetic aperture radar (SAR)~\cite{zheng2024random,zhang2026design} introduces not merely another sensing modality, but a fundamentally different system-level capability. Unlike conventional approaches, which are often constrained by illumination, visibility, and sensing range, SAR provides all-weather, long-range, and illumination-independent observation, enabling reliable perception precisely in the conditions where existing methods degrade.
More importantly, the role of SAR in LAE extends beyond onboard sensing. When deployed across satellites, high-altitude platform stations (HAPS), UAVs, and ground systems, SAR forms a shared environmental sensing layer that provides large-scale and persistent awareness of the physical environment. This layer can supply UAVs with global environmental context, including geometry, occlusion patterns, and dynamic changes, thereby enabling trajectory planning, risk prediction, and coordinated multi-agent operation. In this sense, SAR transforms UAVs from locally reactive agents into globally informed and environment-aware entities.
At the same time, SAR-equipped UAVs directly enhance mission-level capabilities by enabling physics-level sensing tasks that go beyond conventional visual perception. These include infrastructure monitoring, wide-area surveillance, subsurface and hidden-risk detection, and disaster assessment under degraded visibility. As a result, SAR not only supports UAV operation, but also expands what UAVs can fundamentally sense and accomplish.
Therefore, SAR acts as both an enabling infrastructure for environment-aware LAE operation and a capability amplifier for UAV-based applications, jointly driving the transition from fragmented sensing to system-level environmental intelligence.

Building on this perspective, this article highlights a fundamental shift in how sensing should be designed for the LAE. Rather than relying on fragmented and platform-specific sensing, future low-altitude systems require a shared and persistent understanding of the environment.
More broadly, the integration of SAR across satellites, HAPS, UAVs, and ground systems points toward a unified sensing ecosystem, where environmental information is continuously generated, shared, and exploited across platforms. This shift enables environment-aware networking, cooperative operation, and large-scale system optimization, while raising new challenges in deployment and system integration.

Fig.~\ref{fig:overview_transition} illustrates this transformation. On the left, current LAE sensing remains fragmented, local, and highly dependent on environmental conditions, leading to perception gaps under low visibility, occlusion, and wide-area demands. On the right, SAR enables a shared, persistent, and multi-platform sensing layer that provides environment-aware and network-relevant information across the entire airspace. This shift from platform-centric sensing to a shared environmental sensing foundation is central to enabling scalable, reliable, and coordinated LAE systems.

\begin{table*}[t]
\centering
\caption{Comparative capabilities of major sensing modalities for the LAE}
\label{tab:sensing_comparison}
\renewcommand{\arraystretch}{1.2}
\setlength{\tabcolsep}{6pt}
\begin{tabular}{lccccc}
\toprule
\textbf{Capability Dimension} 
& \textbf{Camera} 
& \textbf{LiDAR} 
& \textbf{mmWave Radar} 
& \textbf{Comm.-Assisted Loc./Sens.} 
& \textbf{SAR} \\
\midrule
Low-light / nighttime operation        & Low    & Medium & Medium & N/A    & High \\
Adverse weather robustness             & Low    & Low    & Medium & N/A    & High \\
Occlusion and urban clutter tolerance  & Low    & Low    & Medium & Low    & High$^{\ast}$ \\
Long-range / wide-area monitoring      & Low    & Low    & Low    & Low    & High \\
Non-cooperative target detection       & Low    & Medium & Medium & Low    & High \\
Environmental scene reconstruction     & High   & High   & Low    & Low    & High \\
\bottomrule
\end{tabular}
\vspace{2pt}
\begin{flushleft}
\footnotesize{
High: strong capability; Medium: partial or scenario-dependent capability; Low: limited capability; N/A: not primarily an environmental sensing modality. $^{\ast}$:SAR performance depends on platform geometry, viewing angle, and deployment configuration.
}
\end{flushleft}
\end{table*}

\section{The Missing Link: From Fragmented Sensing to Environment Awareness}

Reliable environmental perception is becoming a critical bottleneck in the evolution of the LAE. While UAV platforms, communication systems, and autonomous control have advanced rapidly, sensing capabilities have not scaled at the same pace. Current approaches, primarily based on optical cameras, LiDAR, millimeter-wave radar, and communication-assisted sensing, remain inherently constrained by environmental conditions, sensing range, and deployment scale~\cite{fan20244d}. Although effective for routine navigation and localized tasks, these modalities struggle to meet the demands of dense, large-scale, and safety-critical operations. This mismatch raises a fundamental question: can existing sensing paradigms truly support the scalability of LAE systems?

\subsection{Why Existing Sensing Fails}

In practice, sensing failures in LAE systems are not rare exceptions, but occur systematically across a wide range of operational scenarios.

\subsubsection{\textbf{Night and Low-Light Conditions}}
Many LAE missions must operate under weak illumination, including nighttime logistics, shaded urban corridors, and emergency scenarios with damaged lighting infrastructure. Optical sensing degrades rapidly under such conditions, while LiDAR suffers from reduced range and sparse point clouds. Millimeter-wave radar can detect coarse structures, but lacks the geometric fidelity required for reliable environmental understanding. As a result, perception becomes increasingly unreliable as illumination weakens.

\subsubsection{\textbf{Rain, Fog, and Smoke}}
Adverse weather and degraded visibility are common in safety-critical LAE missions, such as firefighting, disaster response, and environmental monitoring. Optical imaging becomes ineffective in rain, fog, and smoke, while LiDAR experiences severe attenuation. Although millimeter-wave radar offers improved penetration, it remains insufficient for detailed scene reconstruction. Communication-assisted sensing can only detect device-equipped targets and cannot capture the surrounding environment. Under such conditions, maintaining stable situational awareness is extremely challenging.

\subsubsection{\textbf{Urban Canyons and Complex Terrain}}
Dense urban environments and complex terrain introduce severe occlusion, multipath, and non-line-of-sight conditions. Optical cameras frequently lose visibility, LiDAR struggles to maintain consistent three-dimensional mapping, and communication-assisted sensing becomes biased or unstable. Millimeter-wave radar may detect objects, but often fails to reconstruct the broader environmental structure. These effects result in fragmented and intermittent perception in some of the most common LAE operating environments.

\subsubsection{\textbf{Wide-Area and Long-Range Monitoring}}
As LAE applications expand to infrastructure inspection, cross-regional logistics, and coastal or border surveillance, sensing requirements shift from local perception to wide-area awareness. Optical and LiDAR systems are inherently limited to short ranges, while millimeter-wave radar suffers from resolution degradation with distance. Communication-assisted sensing provides no structural environmental information. Consequently, existing modalities are unable to deliver the wide-swath awareness required for large-scale deployment.

These observations suggest that sensing failures in LAE are not isolated anomalies, but systematic outcomes of fundamental sensing constraints.

\subsection{Why This Is a System-Level Problem}

These recurring failures are not accidental, but arise from deeper structural limitations in how current sensing systems are designed. The fundamental issue is not the weakness of individual sensing modalities, but the fact that their limitations are inherently correlated and often occur simultaneously. Although optical, LiDAR, and radar-based approaches rely on different physical mechanisms, they share common dependencies on visibility, line-of-sight conditions, reflectivity, and sensing range. As a result, the environmental factors that degrade one modality frequently degrade others as well, leading to concurrent sensing failures rather than independent ones.

This leads to a critical mismatch between sensing capabilities and LAE requirements. In real-world missions, challenging conditions such as low visibility, occlusion, and long-range observation demands rarely occur in isolation, but instead co-exist. Consequently, rather than providing complementary information, multi-modal sensing often suffers from correlated perception breakdowns, fundamentally limiting its ability to ensure robust and reliable awareness. Table~\ref{tab:sensing_comparison} summarizes this modality-level mismatch and shows that large portions of low-altitude airspace remain without dependable perception coverage under mission-critical conditions. For completeness, SAR is also included in the table, while its specific role and system relevance will be discussed in the next section.

From a system perspective, current sensing approaches remain inherently local, platform-centric, and environment-dependent. They provide only fragmented and short-range observations, making it fundamentally difficult to construct a consistent and reliable understanding of the broader environment. As a result, UAVs are forced to operate as locally reactive agents, relying on incomplete and unstable observations of their immediate surroundings. More critically, the absence of a global and consistent environmental view prevents proactive decision-making, coordinated multi-agent operation, and scalable system-level optimization.

\subsection{Implication: LAE Cannot Scale Without Shared Environmental Awareness}

The above analysis leads to a fundamental conclusion: current sensing paradigms are insufficient to support the scalability of LAE systems. The challenge is not simply to improve individual sensing modalities, but to overcome the absence of a shared, environment-level perception capability.
Without such a capability, reliable large-scale operation becomes fundamentally unattainable. The lack of persistent and global environmental awareness prevents proactive trajectory planning, coordinated multi-agent operation, and system-level optimization. As a result, UAVs remain constrained to locally reactive behavior, while the overall system fails to achieve the safety, efficiency, and scalability required for large-scale deployment.

This limitation reflects a missing link in current LAE architectures. Sensing must evolve from fragmented, platform-centric observations toward a shared, persistent, and environment-aware perception layer that spans across platforms and scales.
In this context, a fundamentally different sensing paradigm is required. As will be shown in the next section, SAR naturally provides these capabilities, making it a strong candidate for enabling environment-aware and scalable LAE systems.

\begin{figure}[t]
\centering
\includegraphics[width=1.0\linewidth]{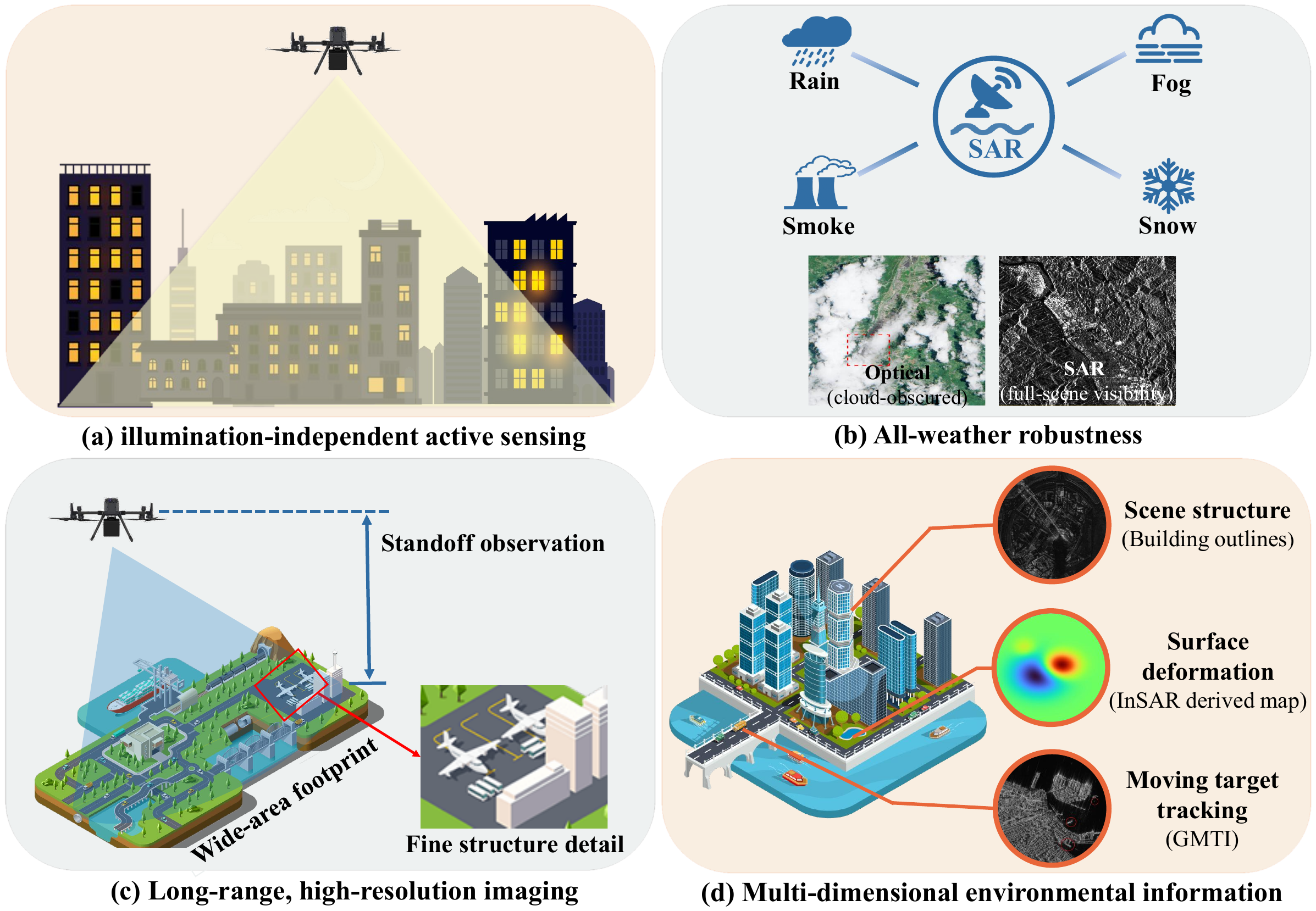}
\caption{Key sensing capabilities of SAR for the LAE.}
\label{fig:sar_capabilities}
\end{figure}
\section{SAR as an Environment-Aware Infrastructure for the LAE}

The previous section established that the sensing limitations in the LAE are not incidental, but structural. Existing sensing modalities fail under correlated conditions, provide only local awareness, and cannot support scalable system-level operation. Addressing these limitations requires a fundamentally different sensing paradigm that can operate reliably across environments, scales, and platforms.
In this context, SAR offers more than incremental improvement. By virtue of its all-weather operation, long-range observation, and environment-level sensing capability, SAR provides a foundation for transforming LAE systems from locally reactive and fragmented sensing toward globally informed and coordinated operation. Importantly, this transformation emerges through three complementary roles of SAR: enabling environment-aware UAV operation, enhancing task-level sensing capabilities, and supporting multi-platform cooperative sensing. Fig.~\ref{fig:sar_capabilities} highlights the key sensing capabilities that underlie this relevance.

\subsection{SAR for UAV: Enabling Environment-Aware Operations}

One of the most fundamental limitations identified in Section II is that UAVs rely on local and incomplete observations, forcing them to operate as locally reactive agents. SAR addresses this limitation by providing large-scale environmental awareness beyond the sensing capability of individual platforms.
When deployed on satellites, high-altitude platform stations (HAPS), or ground systems, SAR can generate wide-area environmental maps that capture structural geometry, occlusion patterns, and large-scale terrain features. These maps provide UAVs with a global environmental context that is otherwise unavailable through onboard sensing alone. As a result, UAVs can perform trajectory planning, obstacle avoidance, and risk assessment based on a broader understanding of the environment rather than immediate local observations.
Moreover, SAR enables environment-aware networking by providing information relevant to communication performance, such as blockage regions, reflective surfaces, and spatial geometry. This supports beam planning, link adaptation, and interference management in wireless-enabled LAE systems. In this sense, SAR serves as a shared environmental awareness layer that connects sensing with communication and control.

Through these capabilities, SAR fundamentally changes the role of UAV sensing, transforming UAVs from blind navigation agents into environment-aware entities capable of predictive and coordinated operation.

\subsection{SAR with UAV: Enabling Task-Level Sensing Capabilities}

Beyond supporting UAV operation, SAR also enhances the sensing capabilities of UAV platforms themselves. Unlike optical and LiDAR-based systems that depend on visibility and illumination, SAR enables sensing based on electromagnetic interaction with the physical environment, allowing UAVs to perform tasks that are fundamentally inaccessible to conventional modalities.

\begin{figure}[t]
\centering
\includegraphics[width=1.0\linewidth]{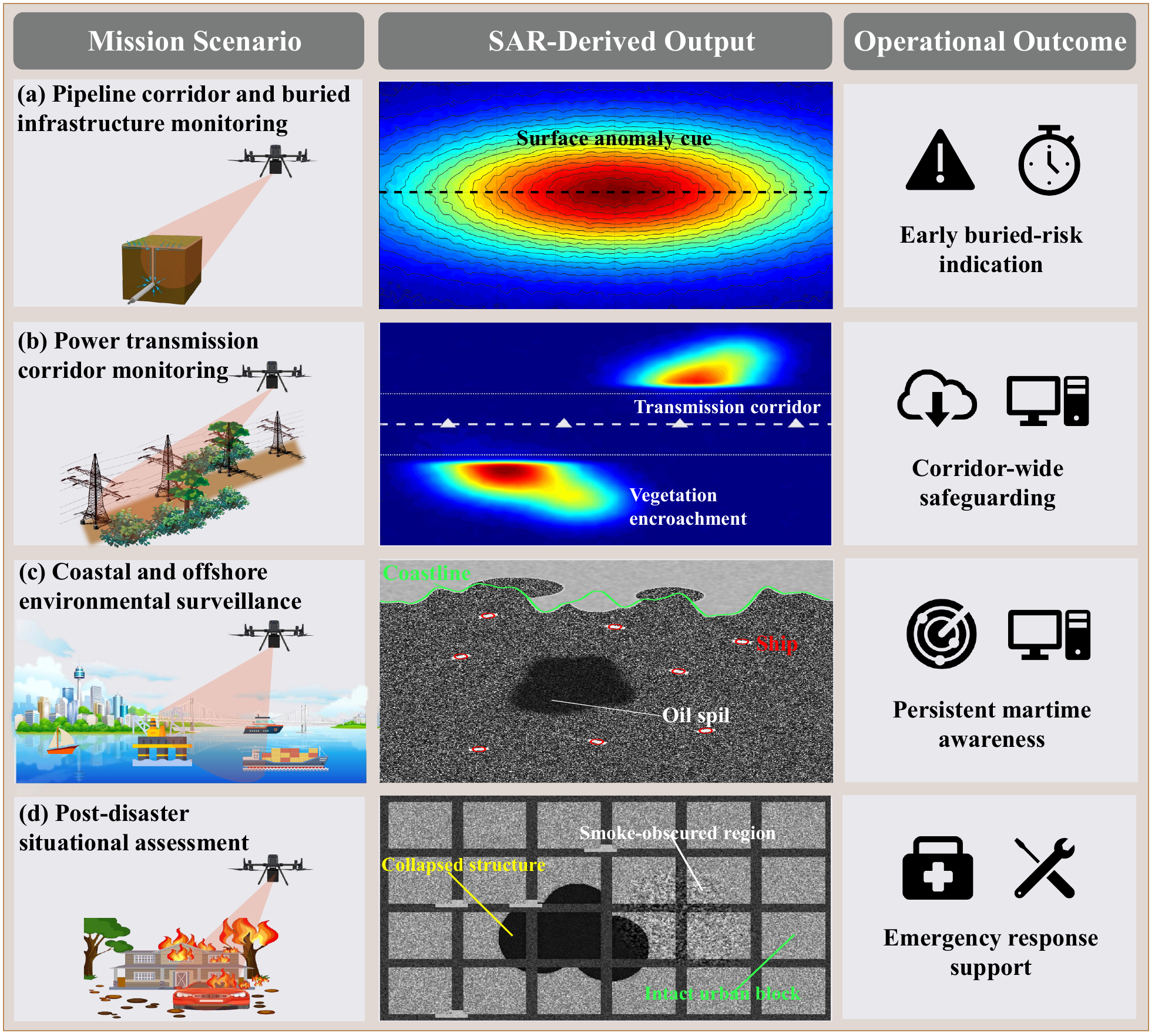}
\caption{Representative SAR-enabled missions in the LAE.}
\label{fig:sar_missions}
\end{figure}

This capability can be broadly categorized into three classes.
First, SAR enables hidden infrastructure sensing, including subsurface or visually obscured objects such as buried pipelines, ground deformation, and infrastructure anomalies. This is critical for applications such as pipeline monitoring, underground risk detection, and infrastructure safety assessment.
Second, SAR supports wide-area monitoring, enabling UAVs to observe large spatial regions such as coastal zones, power transmission corridors, and agricultural fields. This is particularly important for LAE scenarios where coverage, revisit, and scalability are more critical than local detail.
Third, SAR enables perception under degraded visibility, including environments affected by smoke, fog, dust, or low illumination. This is essential for disaster response, emergency assessment, and safety-critical operations where conventional sensing often fails. Hence, the representative mission scenarios are summarized in Fig.~\ref{fig:sar_missions} summarizes several representative SAR-enabled missions and links each mission scenario to its corresponding SAR-derived output and operational outcome.

By enabling these capabilities, SAR extends UAV sensing from vision-based perception to physics-level environmental understanding, significantly expanding what UAVs can detect, interpret, and accomplish.

\subsection{SAR Across Platforms: A Cooperative Sensing Ecosystem}

While SAR provides significant benefits at the individual platform level, its full value in the LAE emerges when it is deployed across multiple platforms and scales. This directly addresses the limitation identified in Section II that no single platform can provide persistent, wide-area, and reliable environmental awareness.
Rather than relying on isolated sensing, SAR enables a cooperative sensing ecosystem in which multiple platforms contribute complementary observations across different spatial and temporal scales. Through such cooperation, environmental information can be continuously generated, updated, and shared, supporting both local sensing and system-level awareness.
This multi-platform perspective highlights the need for a unified sensing framework that spans satellites, aerial platforms, and ground systems. The specific realization of such an architecture will be discussed in the following section.

Such an architecture also aligns naturally with emerging wireless-enabled LAE systems, where sensing, communication, and control must be tightly integrated. SAR-derived environmental information can be distributed across networks, fused with other sensing modalities, and used to support system-level functions such as trajectory optimization, traffic management, and risk prediction.
Through this multi-platform cooperation, SAR enables a scalable and environment-centric sensing paradigm, overcoming the fundamental limitations of fragmented, platform-centric approaches.

\subsection{Summary}

The above discussion highlights that SAR is not simply an additional sensing modality, but a key enabler of system-level transformation in the LAE. By providing environment-aware operation support, task-level sensing enhancement, and multi-platform cooperation, SAR addresses the core limitations identified in Section II.
In particular, SAR mitigates correlated sensing failures through its robustness to illumination and visibility, extends sensing from local to global scales through wide-area observation, and enables coordinated operation through shared environmental awareness.
As a result, SAR enables LAE systems to evolve from fragmented and locally reactive sensing toward globally informed, predictive, and coordinated operation, forming a critical foundation for scalable and reliable low-altitude infrastructure.

\section{System Transformation: How SAR Changes the LAE}

The integration of SAR into the LAE does not merely enhance sensing performance, but fundamentally reshapes how low-altitude systems are structured and operated. By providing persistent, wide-area, and environment-independent perception, SAR introduces a shared and reliable representation of the environment, which enables a transition from fragmented, platform-centric sensing toward infrastructure-level awareness. This shift gives rise to several system-level transformations.

\subsection{From Platform-Centric to Infrastructure-Centric Sensing}

Conventional LAE sensing relies on onboard sensors deployed on individual UAVs, resulting in fragmented and platform-specific perception. Each UAV operates based on its own local observations, leading to inconsistent environmental understanding and redundant sensing efforts across the system.
SAR enables a shift toward infrastructure-centric sensing by decoupling environmental perception from individual platforms. Through deployment across satellites, HAPS, UAVs, and ground systems, SAR provides a shared environmental representation that is persistently available across the operational airspace. As a result, sensing becomes a system-level resource rather than a platform-specific capability. This eliminates redundant sensing tasks, improves consistency in environmental understanding, and enables coordinated operation across multiple agents.

\subsection{From Reactive to Predictive Operation}

Without reliable global awareness, current LAE systems operate in a reactive manner, where UAVs respond only to immediate observations and local conditions.
SAR enables predictive operation by introducing temporal continuity in environmental sensing. Through repeated observations and wide-area coverage, SAR captures not only the current state of the environment, but also its evolution over time. This allows the system to detect structural changes, identify emerging risks, and anticipate future conditions. As a result, UAVs and system controllers can perform proactive trajectory planning, early risk mitigation, and adaptive mission scheduling, significantly improving system robustness and efficiency.

\subsection{From Local Perception to Global Coordination}

A key limitation of existing sensing systems is their inability to support coordinated multi-agent operation at scale. Local perception restricts UAV decision-making to immediate surroundings, making large-scale coordination difficult and often suboptimal.
By providing a global and consistent environmental view, SAR enables coordination across multiple UAVs and platforms. Environmental information becomes a shared reference that allows agents to jointly plan trajectories, avoid conflicts, and allocate tasks based on system-level objectives rather than local heuristics. This capability is essential for dense LAE deployments, where large numbers of UAVs must operate simultaneously in shared airspace.

\subsection{From Device-Centric Networking to Environment-Aware Networking}

In conventional wireless systems, communication and networking are primarily designed based on device locations and channel measurements, without explicit knowledge of the surrounding environment.
SAR enables environment-aware networking by providing explicit information about environmental structures such as obstacles, terrain, and reflective surfaces. This allows communication strategies to be designed with awareness of physical propagation conditions, enabling more effective beamforming, blockage prediction, and trajectory-assisted link optimization. As a result, sensing, communication, and control become tightly coupled, supporting intelligent and adaptive network behavior in LAE systems.

\subsection{Summary}

These transformations reflect a fundamental shift in the role of sensing within the LAE. Rather than serving as a local observation tool, SAR enables sensing to function as a shared infrastructure that supports system-level intelligence.
Through this shift, LAE evolves from a collection of independently operating UAVs into an environment-centric system characterized by shared awareness, predictive operation, and coordinated control. This transition is essential for achieving scalable, reliable, and efficient low-altitude operations.

\begin{figure}[t]
\centering
\includegraphics[width=1.0\linewidth]{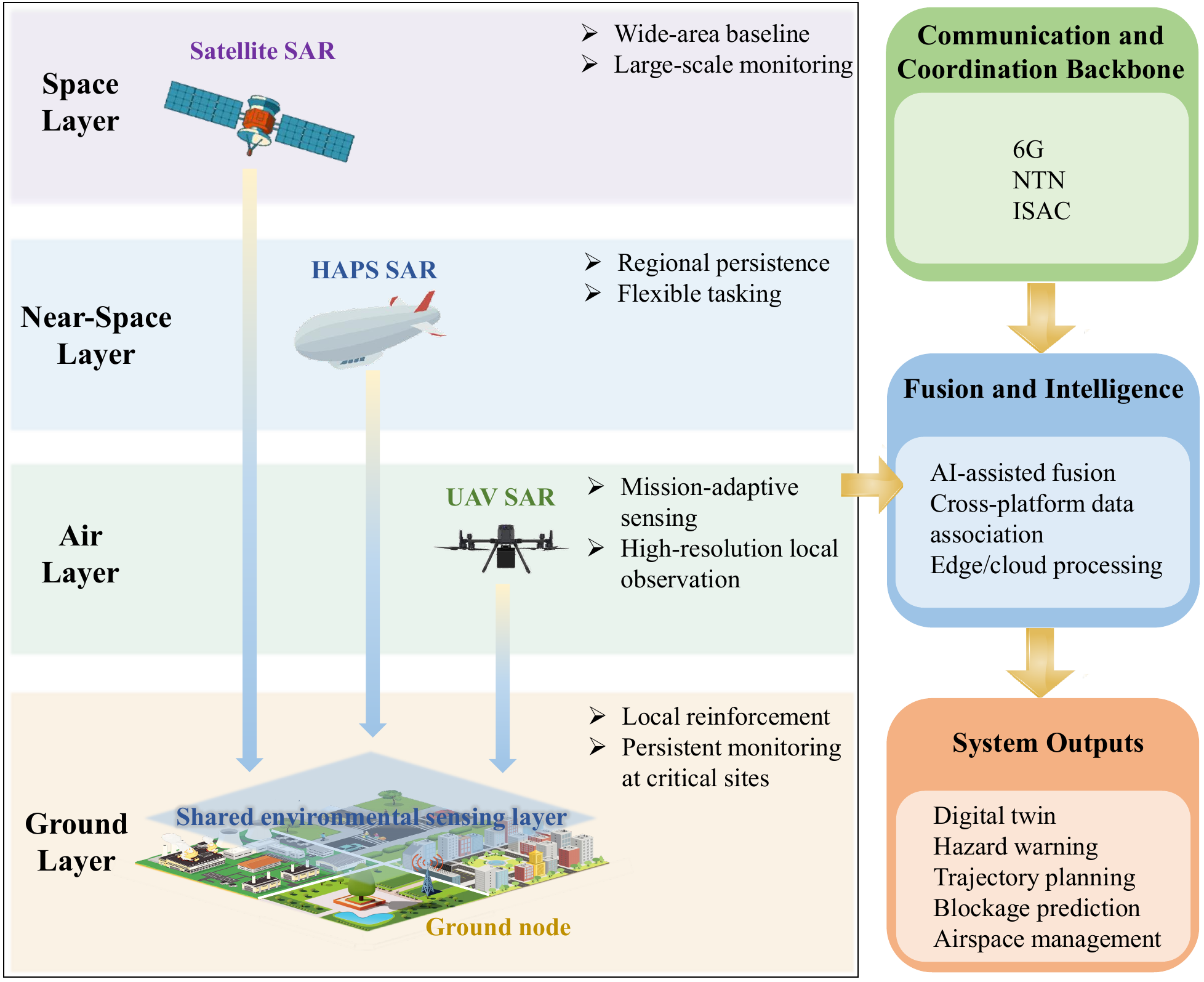}
\caption{Multi-platform SAR-enabled sensing-and-network architecture for the LAE.}
\label{fig:sar_architecture}
\end{figure}  
\section{Enabling Technologies and Architectures}

The system-level transformations discussed in Section IV, including infrastructure-centric sensing, predictive operation, and environment-aware networking, cannot be realized without appropriate technological and architectural support. This section presents the key enablers that allow SAR to function as a shared sensing infrastructure within the LAE.
To realize these transformations, a multi-platform sensing-and-network architecture is required, as illustrated in Fig.~\ref{fig:sar_architecture}.

\subsection{A Multi-Platform SAR-Enabled Sensing Architecture}

A fundamental requirement for environment-aware LAE systems is the ability to provide persistent, wide-area, and multi-scale environmental perception. This cannot be achieved by any single platform alone, but requires a hierarchical sensing architecture spanning multiple layers.

In such an architecture, satellite-based SAR provides global-scale coverage and long-term environmental monitoring~\cite{he2025satellite}. High-altitude platform stations (HAPS) offer regional sensing with flexible deployment and extended dwell time~\cite{toka2024integrating}. UAV-mounted SAR enables high-resolution, localized sensing for task-specific missions, while ground-based radar systems provide continuous monitoring in critical areas.

These layers operate at different spatial and temporal scales, but together form a unified sensing framework. Environmental information is continuously generated, updated, and shared across platforms, enabling a transition from isolated sensing to a persistent and system-wide environmental awareness layer. This architecture directly supports the shift from platform-centric to infrastructure-centric sensing.

\subsection{Integration with ISAC and Environment-Aware Networking}

The effectiveness of SAR in LAE systems is further enhanced through its integration with communication and networking technologies. In particular, the convergence of sensing and communication, often referred to as integrated sensing and communication (ISAC)~\cite{feng2025networked}, provides a natural framework for environment-aware system design.
SAR-derived environmental information can be used to optimize communication strategies by accounting for physical factors such as blockage, reflection, and spatial geometry. This enables environment-aware beamforming, link adaptation, and trajectory optimization, aligning communication performance with environmental conditions.

Through this integration, sensing, communication, and control are no longer treated as separate functions, but become tightly coupled components of a unified system. This supports the transition from device-centric networking toward environment-aware networking, enabling more efficient, reliable, and adaptive LAE operations.

\begin{figure*}[t]
\centering
\includegraphics[width=0.9\linewidth]{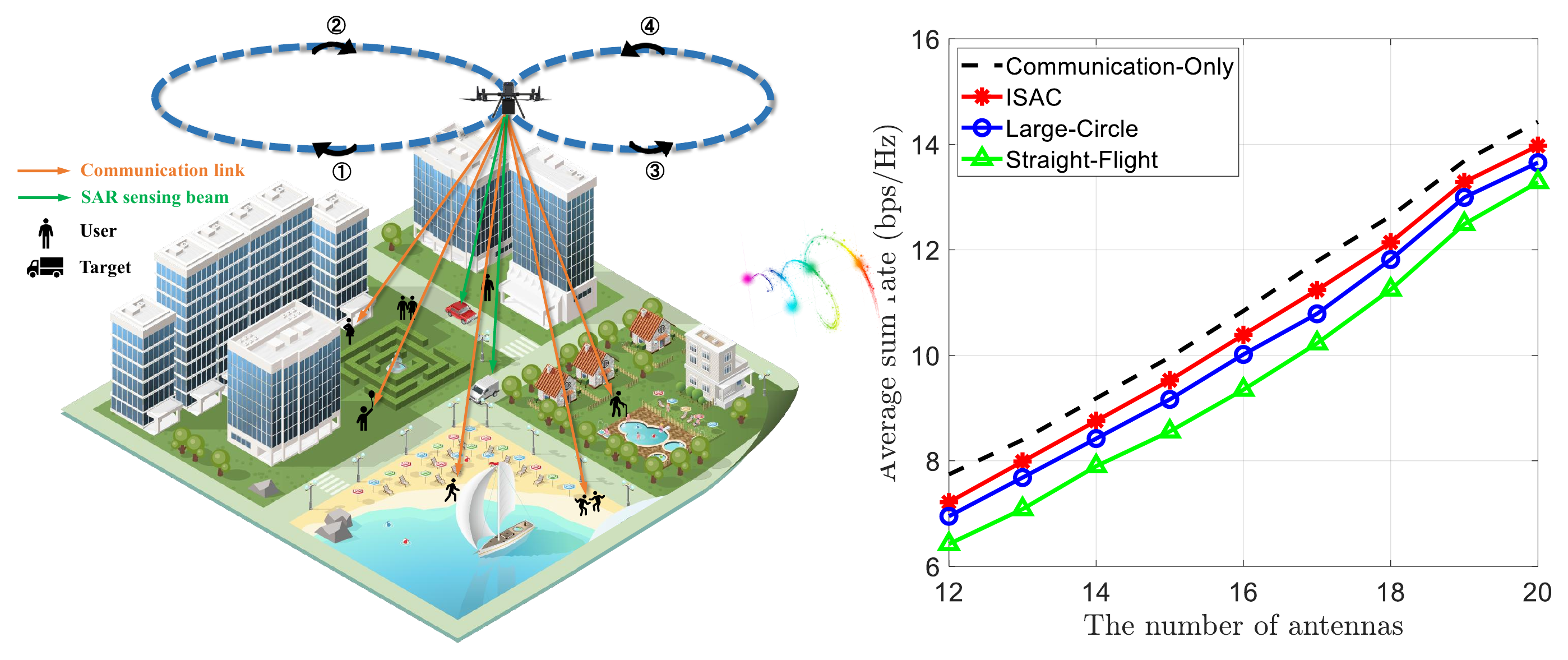}
\caption{A figure-eight UAV-SAR case study for sensing and communication in the LAE.}
\label{fig:figure_eight_isac}
\end{figure*}

\subsection{Case Study: UAV-Based SAR/ISAC with Figure-Eight Trajectory}

To illustrate the practical implications of SAR-enabled LAE systems, consider a UAV-based SAR/ISAC platform employing a figure-eight trajectory, as shown in Fig.~\ref{fig:figure_eight_isac}. In this example, a UAV equipped with a SAR sensing payload simultaneously serves multiple ground users while sensing multiple targets over a mixed low-altitude service region. The considered area includes a denser inner coverage zone and a broader outer footprint, reflecting the heterogeneous spatial demands that arise in practical LAE scenarios.
The UAV follows a closed figure-eight trajectory composed of two connected circular flight segments. This trajectory enables periodic SAR revisit of a shared intersection region while repeatedly observing adjacent subregions, thereby enhancing both sensing persistence and spatial coverage. Compared to conventional circular or linear flight patterns, the figure-eight trajectory provides increased spatial diversity and improved coverage efficiency.

From a sensing perspective, the diversity of observation angles improves imaging quality and robustness, particularly for targets located in complex environments. From a communication perspective, the trajectory introduces more flexible link geometry, which can be exploited for adaptive beamforming, interference management, and improved connectivity.
To highlight the role of trajectory design, the figure-eight operation is compared with three representative baselines: a large-circle trajectory emphasizing broad single-loop coverage, a straight-flight trajectory offering simpler mobility but limited revisit capability, and a communication-only design without SAR sensing. These comparisons suggest that the figure-eight trajectory provides a favorable balance among sensing persistence, observation diversity, and communication performance.
More importantly, this case study demonstrates how trajectory design, sensing, and communication can be jointly optimized within an environment-aware framework. It highlights the potential of SAR-enabled UAV systems to simultaneously support perception and communication objectives, reinforcing the concept of integrated system-level design in LAE systems.

\subsection{Toward Intelligent LAE Systems: AI and Digital Twins}

As LAE systems grow in scale and complexity, the role of data-driven intelligence becomes increasingly important. The large-scale environmental data generated by SAR systems provides a rich foundation for artificial intelligence (AI)~\cite{zhao2025generative} and digital twin~\cite{he2025digital} technologies.
By integrating SAR data into digital twin models of the environment, it becomes possible to create dynamic and predictive representations of the physical world. These models can support real-time decision-making, risk assessment, and system optimization across multiple platforms.
AI techniques can further enhance this process by enabling data fusion, anomaly detection, and predictive analytics. Together, SAR, AI, and digital twin technologies enable the development of intelligent LAE systems that are not only environment-aware, but also adaptive and self-optimizing.
Through these enabling technologies and architectures, the system-level transformations introduced by SAR can be effectively realized in practical LAE deployments.

\section{Challenges Toward Real Deployment}

Despite its potential to enable environment-aware and scalable LAE systems, the practical deployment of SAR-based sensing architectures faces several significant challenges. These challenges are not limited to individual technologies, but arise from the need to integrate sensing, communication, and control across multiple platforms and scales.

\subsection{System Integration and Cross-Layer Coordination}

One of the primary challenges lies in integrating SAR sensing with UAV operation, communication systems, and control mechanisms. Unlike conventional sensing modalities, SAR operates at multiple spatial and temporal scales, often across heterogeneous platforms such as satellites, HAPS, UAVs, and ground systems.
Coordinating these platforms requires cross-layer design that jointly considers sensing objectives, communication constraints, and trajectory planning. Achieving such integration is challenging due to the differing requirements of each subsystem, including latency, data rate, sensing resolution, and energy consumption. Ensuring consistent and reliable operation across these layers remains an open problem.

\subsection{Scalability and Deployment Cost}

The deployment of SAR systems across multiple platforms introduces significant challenges in scalability and cost. Satellite and HAPS-based SAR systems require substantial infrastructure investment, while UAV-mounted SAR systems are constrained by payload, power consumption, and flight endurance.
As LAE systems scale to support dense and widespread operations, the cost and complexity of deploying and maintaining a multi-platform SAR infrastructure become critical considerations. Balancing performance with deployment feasibility is essential for practical implementation.

\subsection{Real-Time Processing and Data Fusion}

SAR systems generate large volumes of high-resolution data, which poses challenges for real-time processing and integration into operational decision-making. Unlike conventional sensing data, SAR data often requires computationally intensive processing for imaging, interpretation, and fusion with other modalities.
Enabling real-time or near-real-time utilization of SAR-derived information requires advances in efficient signal processing, edge computing, and distributed data fusion. Integrating SAR data with other sensing modalities and communication systems further increases the complexity of the processing pipeline.

\subsection{Standardization, Regulation, and Airspace Governance}

Beyond technical challenges, the deployment of SAR-enabled LAE systems must address regulatory and governance issues. The use of multiple sensing platforms operating across shared airspace raises concerns related to spectrum allocation, data privacy, and operational safety.
In addition, the lack of standardized frameworks for integrating sensing, communication, and control across platforms limits interoperability and large-scale adoption. Developing regulatory policies and standardization efforts that support environment-aware LAE systems is essential for enabling practical deployment.

\subsection{Summary}

Addressing these challenges is critical for translating the system-level potential of SAR into real-world LAE applications. Overcoming these barriers will require coordinated advances in system design, algorithm development, infrastructure deployment, and policy frameworks.
\section{Conclusion}

The rapid evolution of the low-altitude economy is fundamentally reshaping how aerial systems are deployed and operated. As this paper has shown, the primary challenge is no longer enabling UAV flight, but establishing reliable and scalable environmental awareness across complex and dynamic airspace.
In this context, SAR represents a key enabler of a system-level transformation. By providing persistent, wide-area, and environment-independent sensing, SAR addresses the fundamental limitations of existing modalities and enables a shift from fragmented, platform-centric perception to a shared environmental sensing infrastructure.
Through this transformation, LAE systems evolve from locally reactive and isolated operations toward globally informed, predictive, and coordinated systems. This shift not only enhances sensing capabilities, but also enables tighter integration between sensing, communication, and control, forming the foundation of intelligent and adaptive low-altitude networks.
Looking forward, the integration of SAR with emerging technologies such as ISAC, artificial intelligence, and digital twins is expected to further accelerate this evolution. Realizing this vision will require continued advances in system integration, scalable architectures, and regulatory frameworks, paving the way toward environment-centric LAE systems.
In this sense, SAR enables the LAE to evolve from flight-centric systems into environment-aware infrastructure, supporting the next generation of scalable, reliable, and intelligent low-altitude operations.

\bibliographystyle{IEEEtran}
\bibliography{references}

%\section*{Biographies}
%\begin{IEEEbiographynophoto}
%{Xue Zhang} received the B.S. degree from Southwest University (SWU), China, in 2020, and the M.S. degree from the University of Electronic Science and Technology of China (UESTC), in 2023. She is currently pursuing the Ph.D. degree with King Abdullah University of Science and Technology (KAUST), Thuwal, Saudi Arabia.
%\end{IEEEbiographynophoto}

%\begin{IEEEbiographynophoto}
%{Bang Huang} 
%\end{IEEEbiographynophoto}

%\begin{IEEEbiographynophoto}
%{Mohamed-Slim Alouini} [S’94, M’98, SM’03, F’09] received his Ph.D. degree in electrical engineering from the California Institute of Technology, Pasadena, in 1998. He served as a faculty member at the University of Minnesota, Minneapolis, then at Texas A$\&$M University at Qatar, Doha, before joining KAUST as a professor of electrical engineering in 2009. His current research interests include the modeling, design, and performance analysis of wireless communication systems.
%\end{IEEEbiographynophoto}

\end{document}